\documentclass[aps,prl,showpacs,twocolumn,groupedaddress]{revtex4}

\usepackage{graphicx}
\usepackage{units}
\usepackage{amsmath}
\begin{document}
\title{Algebraic fidelity decay for local perturbations}
\author{R. H\"{o}hmann}
\author{U.~Kuhl}
\author{H.-J.~St\"{o}ckmann}
\affiliation{Fachbereich Physik, Philipps-Universit\"{a}t Marburg,
Renthof 5, D-35032 Marburg, Germany}
\date{\today}

\begin{abstract}
From a reflection measurement in a rectangular microwave billiard with randomly distributed scatterers
the scattering and the ordinary fidelity was studied. The position of one of the scatterers is the perturbation
parameter. Such perturbations can be considered as {\em local} since wave functions are influenced only locally,
in contrast to, e.\,g., the situation where the fidelity decay is caused by the shift of one billiard wall. Using
the random-plane-wave conjecture, an analytic expression for the fidelity decay due to the shift of one
scatterer has been obtained, yielding an algebraic $1/t$ decay for long times. A perfect agreement between
experiment and theory has been found, including a predicted scaling behavior concerning the dependence
of the fidelity decay on the shift distance. The only free parameter has been determined independently
from the variance of the level velocities.
\end{abstract}


\maketitle

A versatile tool to describe the quantum mechanical stability of a
system against perturbations is fidelity introduced by Peres 25
years ago \cite{per84}. The fidelity amplitude is defined as the
overlap integral of an initial state $|\psi_0\rangle$ with itself,
after having experienced two slightly different time evolutions,
\begin{equation}\label{eq::def}
  f(t)=\langle\psi_0|e^{2\pi\imath H_0 t}e^{-2\pi\imath H_1 t}|\psi_0\rangle.
\end{equation}
Here $H_0$ and $H_1=H_0+\lambda V$ are the two relevant
Hamiltonians, and parameter $\lambda$ describes the strength of the
perturbation. Throughout this letter the time will be given in units
of the Heisenberg time $t_H= \hbar/\Delta E$, where $\Delta E$ is
the mean level spacing. The fidelity $F(t)$ is the modulus square of
the fidelity amplitude. The concept of fidelity is closely related
to the concept of decoherence which is widely used in the community
of quantum computation \cite{kar02,gor04b}.

For very weak perturbation strengths $\lambda$, in the
perturbative regime, Eq.~(\ref{eq::def}) reduces to
\begin{equation}\label{eq::pert}
  f(t)=\langle\psi_0|e^{-2\pi\imath \lambda V_{\rm diag} t}|\psi_0\rangle\,,
\end{equation}
where $V_{\rm diag}$ is the diagonal part of the perturbation in the
basis of $H_0$. Assuming a Gaussian distribution for the matrix
elements of $V_{\rm diag}$ Eq.~(\ref{eq::pert}) describes a Gaussian
decay of the fidelity beyond the Heisenberg time. With increasing
perturbation strength the Gaussian decay is taken over by an
exponential decay below the Heisenberg time with a decay constant
which can be calculated from the matrix elements of the perturbation
by means of Fermi's golden rule \cite{jal01,jac01b,cer02}. For
perturbations with missing diagonals the Gaussian decay is
suppressed leaving only the exponential decay, a situation termed
freeze \cite{pro05}. For very strong perturbation strengths the
fidelity decay becomes independent on the perturbation strength and
reflects the classical dynamics. In chaotic systems  an exponential
decay with the classical Lyapunov exponent is observed \cite{cuc02},
whereas in integrable and diffusive systems an algebraic tail is found
\cite{jac03,ada03}. For more details it is referred to the review
article by Gorin et al \cite{gor06c}.

In all references mentioned above {\em global} perturbations have
been considered, meaning that there is a total rearrangement of
spectrum and eigenfunctions already for moderate perturbation
strengths. For this situation random matrix theory, assuming
uncorrelated Gaussian distributions for the matrix elements of $V$,
has yielded a number of perturbative  and exact results
\cite{gor04a,stoe05}. The fidelity decay by global perturbations has
been studied experimentally in a microwave billiard, where one wall
was shifted \cite{sch05b}, and in a vibrating aluminum block, where
the temperature took the role of the perturbation parameter
\cite{lob03b,gor06b}. In chaotic systems the long time fidelity
for global perturbation always is either Gaussian or single exponential,
a very unsatisfactory situation, e.\,g.,  for quantum computing.

The role of {\em local} perturbations, realized in billiard systems,
e.\,g., by the shift of an impurity, has been more or less ignored
in the past. This is somewhat surprising, since many, if not most of
the perturbing interactions in real systems are  short ranged.
Examples are diffusive jumps or flips of neighbored spins in solids
leading to the decay of spin echoes in nuclear magnetic resonance
(see  e.\,g.~Ref.~\cite{sli80}). Another example is the twinkling of
stars, caused by the diffusive motion of the atoms in the
atmosphere.

In a previous work we studied the spectral level dynamics in
microwave billiards for global variations, realized by a shift of
one wall, and local variations, realized by the shift of an impurity
\cite{bar99d}. For global variations a Gaussian eigenvalue velocity
distribution was found, as expected for chaotic systems. For local
perturbations the situation was found to be completely different.
The introduction of a point-like impurity leads to a shift of the
eigenvalues by
\begin{equation} \label{eq::EnergyShift}
  \Delta E_n = \alpha|\psi_n(r)|^2\,,
\end{equation}
where $\alpha$  describes the strength of the perturbation, and
$\psi_n$ is the  wave function at the scatterer position in the
absence of the scatterer. Using the random-plane-wave assumption a
modified Bessel function was obtained for the velocity distribution.
With increasing radius of the impurity or, equivalently, with
increasing frequency  a gradual transition from local to global
behavior was found. Somewhat later our results had been verified in
supersymmetry calculations \cite{mar03}.

In view of the fact that the velocity distributions for global and
local perturbations are completely different one could expect that
the same is true for the fidelity decay. It will be shown in this
letter that this is really the case. We shall see that for local
perturbations the fidelity decays algebraically for long times,
i.\,e. the situation is much more favorable  than it is the case for
global perturbations.

The fidelity in its original definition (\ref{eq::def}) is hardly
accessible experimentally. This was our motivation in a previous
work \cite{sch05b} to introduce the scattering fidelity being
defined as
\begin{equation}\label{eq::scatfid}
f^{\mathrm{scat}}(t)= \frac{\displaystyle\hat
C[S_{ab},S_{ab}^{(\lambda)\ast}](t)}{\displaystyle \sqrt{ \hat
C[S_{ab},S_{ab}^{\ast}](t) \cdot \hat
C[S_{ab}^{(\lambda)},S_{ab}^{(\lambda)\ast}](t)}}\,.
\end{equation}
Here $S_{ab}$ is a scattering matrix element, and $\hat
C[S_{ab},S_{ab}^{(\lambda)\ast}](t)$ the Fourier transform of the
cross-correlation function of the respective matrix elements.
It was stated in Ref.~\cite{sch05b},
that for chaotic systems and weak coupling of the measuring antenna
the scattering fidelity approaches the ordinary fidelity. In the
present work it has been possible to study both quantities at the
same time and to compare the results. Since in all previous
experiments on  fidelities actually scattering fidelities or related
quantities had been measured, though without stating this
explicitly, such a comparison is of fundamental interest.

\begin{figure}
  \includegraphics[width=.9\columnwidth]{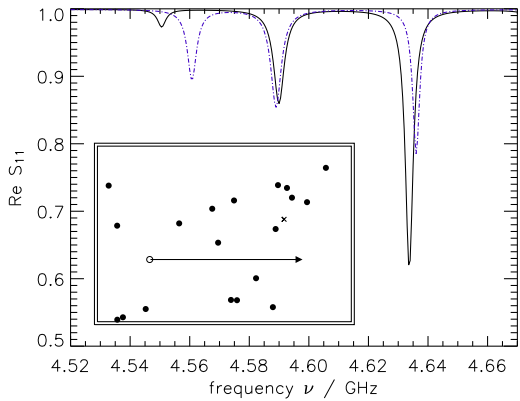}
\caption{\label{fg::expSketch}
Part of the reflection spectrum for
two perturber positions differing by $|\Delta r|=\unit[4]{mm}$. The
inset shows a sketch of the used microwave billiard (scatterers are
shown three times larger). The antenna position is marked by a
cross.}
\end{figure}

The system  studied was a rectangular microwave billiard with height
$h=\unit[8]{mm}$, side lengths $a=\unit[340]{mm}$,
$b=\unit[240]{mm}$ and 19 disks with a diameter of
$D=\unit[4.6]{mm}$ placed randomly inside the billiard (see inset of
Fig.~\ref{fg::expSketch}). Another disk of the same size has been
moved in steps of $|\Delta r| =\unit[1]{mm}$ through the billiard.
With a fixed wire antenna we measured the reflection spectrum for
300 different positions of the moving disk in a frequency range from
3.5\,GHz to 6\,GHz. In this frequency range the billiard can still
be treated as two-dimensionally, and there is a complete agreement
with the corresponding quantum mechanical system  \cite{stoe99}.
Figure~\ref{fg::expSketch} shows part of the reflection spectrum for
two slightly different positions of the movable disk. In addition we
measured the eigenfunctions \cite{ste92} to make sure that they are
delocalized and their intensities Porter-Thomas distributed.
Thus the system is fully chaotic and ballistic in the investigated frequency range.
In the measured frequency regime it was also possible to extract resonance
positions and amplitudes from the spectrum by a Lorentz fit. These
quantities will be needed later for the determination of the
ordinary fidelity.

First we shall derive an explicit expression of the fidelity decay
caused by the shift of a local perturber. We assume that the
system is completely chaotic (in the experiment this was verified by
measuring the eigenfunctions, see above). In this case we may
average expression (\ref{eq::def}) over all possible initial states
resulting in
\begin{equation}\label{eq::DefAvFid}
f(t) = \frac{1}{N}{\rm Tr}\left(e^{2\pi\imath H_0 t}e^{-2\pi\imath
H_1 t}\right)\,,
\end{equation}
where $N$ is the number of states taken in the trace. In the present
case $H_0$ and $H_1$ correspond to the Hamiltonian of the billiard
with the perturber placed at two different positions. For a weak,
point-like perturbation, the perturber just produces a small 
shift of the eigenenergy proportional to the intensity $|\psi|^2$ of
the unperturbed wave function at the perturber position,
see Eq.~(\ref{eq::EnergyShift}), i.\,e. the Hamiltonian in the basis
of the unperturbed system is given by
\begin{equation}\label{eq::HamiltonOp}
H_{nm}(r) = \delta_{nm}(E_n^0+\alpha|\psi_n(r)|^2)\,,
\end{equation}
where the $E_n^0$ are the eigenenergies of the unperturbed system.
For this approach to be valid it is essential that the shift induced
by the perturber is always small compared to the mean level spacing,
i.\,e. we never leave the perturbative regime. We then obtain from
Eqs.~(\ref{eq::pert}) and (\ref{eq::HamiltonOp})
\begin{equation}
f(t) = \left\langle
e^{2\pi\imath\alpha(|\psi_1|^2-|\psi_2|^2)}\right\rangle\,,
\end{equation}
where the abbreviation $\psi_i=\psi(r_i)$ has been used. The
brackets denotes the average over all $N$ participating states. To
calculate this average, we now apply the random plane wave
assumption. The average is most conveniently  expressed in the
following way \cite{sre96b},
\begin{eqnarray}
f(t) &=& \frac{\sqrt{|K|}}{2\pi}\int\!\!\int d\psi_1d\psi_2 e^{
2\pi\imath\alpha(|\psi_1|^2-|\psi_2|^2)}\nonumber\\
&&\qquad\qquad\quad\times e^{-\frac{1}{2}(\psi_1, \psi_2)K(\psi_1,
\psi_2)^T}\,,
\end{eqnarray}
where $K$ is a $(2\times2)$-matrix of which the inverse
\begin{equation}
    K^{-1} = \left(\begin{aligned} \langle\psi_1\psi_1\rangle& \quad & \langle\psi_1\psi_2\rangle \\
        \langle\psi_2\psi_1\rangle& \quad & \langle\psi_2\psi_2\rangle\\
        \end{aligned}\right)
\end{equation}
can be expressed in terms of two-point correlation functions only.
Using again the model of random plane waves \cite{ber77a} the
two-point correlation function can be calculated yielding
$\langle\psi_i\psi_j\rangle=\frac{1}{A}J_0(k|r_i-r_j|)$, where $A$
is the billiard area. Performing the integrations one obtains the
final expression for the fidelity amplitude
\begin{equation}\label{eq::FidelityResult}
  f(t) = \left[1+(\lambda t)^2\right]^{-\frac{1}{2}}\,,
\end{equation}
where
\begin{equation}\label{eq::lambda}
  \lambda = \frac{4\pi\alpha}{A}\sqrt{1-J^2_0(k |\Delta r|)}\,.
\end{equation}
\begin{figure}
  \includegraphics[width=\columnwidth]{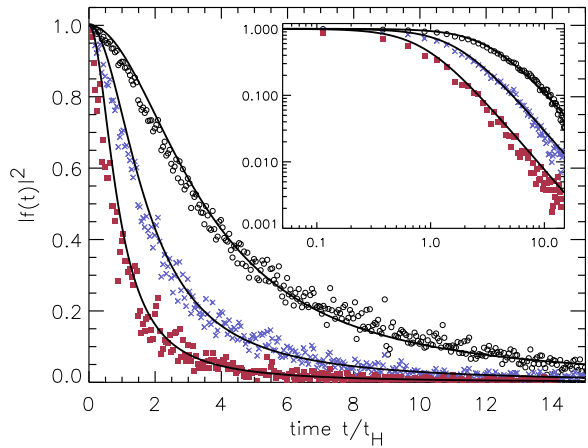}
\caption{\label{fg::ExpScatFid}
Scattering fidelity (color online)
for different shifts of the perturber $|\Delta r|=1$\,mm (open
circles, black), 2\,mm (crosses, blue), and 4\,mm (filled squares,
red). The solid line corresponds to the theoretical prediction of
Eq.~(\ref{eq::FidelityResult}). The results were obtained for the
frequency range from 3.5\,GHz to 6\,GHz. The insert shows the same
results in a double-logarithmic plot.}
\end{figure}
\begin{figure}
  \includegraphics[width=\columnwidth]{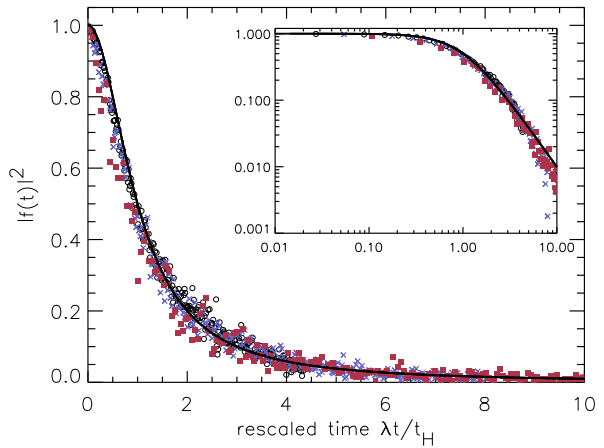}
\caption{\label{fg::ExpScatScaledFid}
Same as Fig.~\ref{fg::ExpScatFid},
but plotted on a rescaled time axis using
the scaling property of Eq.~(\ref{eq::FidelityResult}).}
\end{figure}%
$|\Delta r|$ is the shift of the scatterer. For large $t$ the
fidelity amplitude decays algebraically with $f(t)\sim 1/t$. The
only free parameter in Eq.~(\ref{eq::lambda}) is $\alpha$. This
parameter can be obtained independently from the variance of the
level velocities (see Ref. \onlinecite{cer02}). This allows us to
compare the experimental results with theory without any free
parameter.

Figure~\ref{fg::ExpScatFid} shows the scattering fidelity, as
obtained from Eq.~(\ref{eq::scatfid}), for three different perturber
shifts. The solid lines correspond to the theoretical prediction
from Eq.~(\ref{eq::FidelityResult}). A good agreement between theory
and  experiment is found in all cases, apart from minor systematic
deviations for small times. Values obtained for the coupling
parameter $\alpha$ by a fit deviate from those from the level
velocities by a few \%. The inset shows the results in a
double-logarithmic plot illustrating  the algebraic decay for long
times $t$. Equation~(\ref{eq::FidelityResult}) exhibits  a scaling
behavior: On a rescaled time axis $\lambda t$  all experimental
results should fall onto one single curve.
Figure~\ref{fg::ExpScatScaledFid} demonstrates that this is indeed
the case.

So far we have discussed the results for the scattering fidelity.
Let us now see, how the ordinary fidelity (\ref{eq::def}) can be
obtained. Since the measurement has been performed at a fixed
antenna position $r_0$, the initial state is localized at $r_0$,
i.\,e. $|\psi_0\rangle=|r_0\rangle=\delta(r-r_0)$. Expanding
$|\psi_0\rangle$ in terms of eigenfunctions $|\psi_n^1\rangle$ of
$H_1$, and $\langle \psi_0|$ in terms of eigenfunctions $\langle
\psi_n^0|$ of $H_0$, Eq.~(\ref{eq::def}) reads
\begin{equation}
    f(t)=\sum\limits_{n,m}\psi_n^0(r_0)\psi_m^1(r_0)
    e^{2\pi\imath\left(E_n^0-E_m^1\right)t}\langle\psi_n^0|\psi_m^1\rangle\,.
\end{equation}
The  shift of the energy is a first order effect of the perturber,
the change of the eigenfunctions being of the next order. Neglecting
these higher order effects, we may approximate
$\langle\psi_n^0|\psi_m^1\rangle\approx \delta_{nm}$ to obtain
\begin{equation}\label{eq::resofid}
  f(t)=\sum\limits_n\psi_n^0(r_0)\psi_n^1(r_0)
  e^{2\pi\imath\left(E_n^0-E_n^1\right)t}\,.
\end{equation}
\begin{figure}
  \includegraphics[width=\columnwidth]{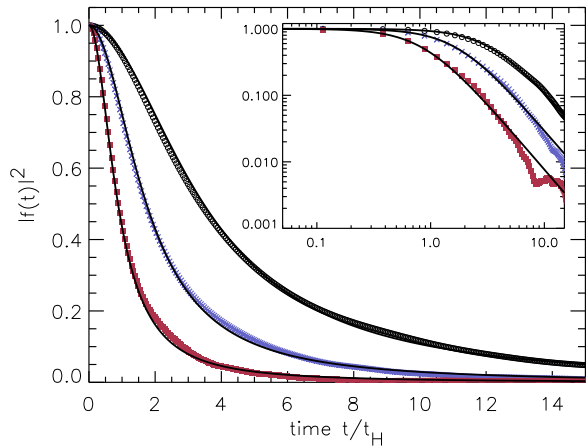}
\caption{\label{fg::ExpDirFid}
Ordinary fidelity determined as
obtained from the resonance position and depths, see text. The
meaning of the symbols is the same  as in Fig.~\ref{fg::ExpScatFid}.
}
\end{figure}
\begin{figure}
  \includegraphics[width=\columnwidth]{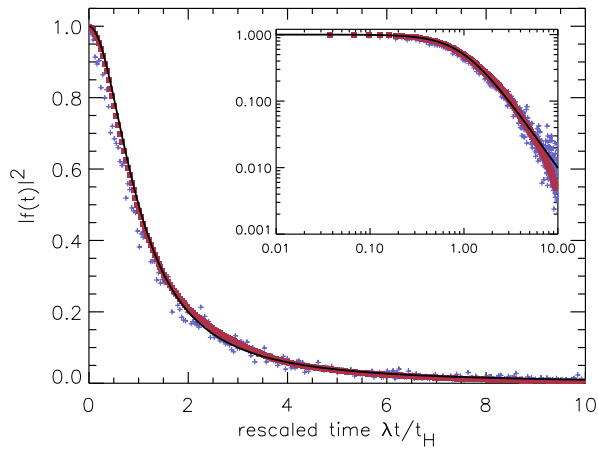}
\caption{\label{fg::ExpAllFid}
Scattering fidelity, obtained from a
superposition of the results from all perturber shifts (blue
crosses) and ordinary fidelity (red squares)  on a rescaled time
axis.  The solid line corresponds to the  theoretical prediction.}
\end{figure}
All quantities entering the sum on the right hand side are available
from the experiment, the eigenenergies from the resonance positions,
and the eigenfunctions at the antenna position from the resonance
depths \cite{kuh00b}. Altogether 64 resonances have been taken. In
Fig.~\ref{fg::ExpDirFid} the results for the fidelity are presented.
A very good agreement between theory and experiment is found. Here
the relative deviation for the coupling parameter $\alpha$ taken from
the variance of the level velocities and the one obtained from the
fidelity even agree up to 1\%. No smoothing has been applied.
The ordinary fidelity, too, shows
the scaling behavior predicted by Eq.~(\ref{eq::FidelityResult}) as
is pronounced in Fig.~\ref{fg::ExpAllFid}. In addition the collected
results for the scattering fidelity, obtained by an averaging over
all perturber shifts, is plotted, showing the quality of the
agreement between both types of fidelity.

In summary we have shown that  for a local perturbation, realized by
the shift of a perturber in a microwave billiard, the fidelity
decays algebraically with a $1/t$ long-time behavior. This is in
contrast to  the exponential or Gaussian long-time behavior observed
in the fidelity decay due to global perturbations. All results could
be quantitatively explained within the random plane wave model,
including a scaling prediction on the dependence of the fidelity
decay on the perturber shift. In addition we could show that
scattering fidelity and ordinary fidelity are identical within the
experimental errors.

T. Seligman, Cuernavaca, Mexico, is thanked for numerous discussions
and suggestions. This work was founded by the Deutsche
Forschungsgemeinschaft via the Forschergruppe 760 ``Scattering
Systems with Complex Dynamics''.

\end{document}